\documentclass[aps,10pt,notitlepage,twocolumn, superscriptaddress,nofootinbib,longbibliography]{revtex4-2}

\usepackage[latin1]{inputenc}
\usepackage[T1]{fontenc}
\usepackage{lmodern}
\usepackage{amsmath}
\usepackage{amssymb}
\usepackage{graphicx}
\usepackage{float}
\usepackage{esint}
\usepackage{longtable}
\usepackage{dcolumn}
\usepackage{babel}
\usepackage{csquotes}
\usepackage{color}
\usepackage{slashed}
\usepackage{simplewick}
\usepackage{amsmath,latexsym}

\usepackage{cancel}

\usepackage{hyperref}
\hypersetup{
    colorlinks,
    citecolor=blue,
    filecolor=green,
    linkcolor=purple,
    urlcolor=red,
}

\usepackage{slashed}

\usepackage{hyperref}
\hypersetup{colorlinks,breaklinks,
			citecolor=[rgb]{0,0.0,1.0},
            urlcolor=[rgb]{0.0,0.0,1.0},
            linkcolor=[rgb]{0,0.5,0.9}}

\begin{document}

\title{Nonsingular, lump-like, scalar compact objects in $(2+1)$-dimensional Einstein gravity}

\author{Roberto V. Maluf}
\email{r.v.maluf@fisica.ufc.br}
\affiliation{Universidade Federal do Cear\'a (UFC), Departamento de F\'isica,\\ Campus do Pici, Fortaleza - CE, C.P. 6030, 60455-760 - Brazil.}
\affiliation{Departamento de F\'{i}sica Te\'{o}rica and IFIC, Centro Mixto Universidad de Valencia - CSIC. Universidad
de Valencia, Burjassot-46100, Valencia, Spain.}

\author{Gerardo Mora-P\'{e}rez}
\email{moge@alumni.uv.es}
\affiliation{Departamento de F\'{i}sica Te\'{o}rica and IFIC, Centro Mixto Universidad de Valencia - CSIC. Universidad
de Valencia, Burjassot-46100, Valencia, Spain.}

\author{Gonzalo J. Olmo}
\email{gonzalo.olmo@uv.es}
\affiliation{Departamento de F\'{i}sica Te\'{o}rica and IFIC, Centro Mixto Universidad de Valencia - CSIC. Universidad
de Valencia, Burjassot-46100, Valencia, Spain.}
\affiliation{Universidade Federal do Cear\'a (UFC), Departamento de F\'isica,\\ Campus do Pici, Fortaleza - CE, C.P. 6030, 60455-760 - Brazil.}

\author{Diego Rubiera-Garcia}
\email{drubiera@ucm.es}
\affiliation{Departamento de F\'{i}sica T\'{e}orica and IPARCOS,\\ Universidad Complutense de Madrid, E-28040 Madrid, Spain.}


\date{\today}

\begin{abstract}
We study the space-time geometry generated by coupling a free scalar field with a non-canonical kinetic term to General Relativity in $(2+1)$ dimensions. After identifying a family of scalar Lagrangians that yield exact analytical solutions in static and circularly symmetric scenarios, we classify the various types of solutions and focus on a branch that yields asymptotically flat geometries. We show that the solutions within such a branch can be divided in two types, namely, naked singularities and nonsingular objects without a center. In the latter, the energy density is localized around a maximum and vanishes only at infinity and at an inner boundary. This boundary has vanishing curvatures and cannot be reached by any time-like or null geodesic in finite affine time. This allows us to consistently interpret such solutions as nonsingular, lump-like, static compact scalar objects, whose eventual extension to the $(3+1)$-dimensional context could provide structures of astrophysical interest.
\end{abstract}


\maketitle

\section{Introduction}

The search for exact analytical solutions to Einstein equations in the presence of reasonable matter sources and couplings is generally a challenging problem, particularly if one works in a $(3+1)$ dimensional space-time. Among the known exact solutions of General Relativity (GR), only a handful are imbued with relevant physical meaning \cite{Stephani:2003tm}. Stationary solutions with spherical and axial symmetries are notable within this group. Indeed, the uniqueness theorems \cite{Penrose:1969pc, Carter:1971zc} assure us that the most general possible asymptotically-flat solution of an electromagnetic nature in vacuum is given by the Kerr-Newman one, which can be interpreted as the gravitational field external to a body described solely by its mass, electric charge, and angular momentum \cite{Kerr:1963ud, Newman:1965my}. Since charge is typically neglected in astrophysical scenarios, the above solution reduces to the Kerr one.

The need to confront observations with the Kerr hypothesis, namely, that all rotating, fully-collapsed, objects in the Universe belong to the Kerr family of solutions, has sparked much interest in recent years to obtain new solutions that describe alternative compact astrophysical objects. Solutions of that type are not interesting only for their observational characteristics, such as shadows \cite{EventHorizonTelescope:2019dse,EventHorizonTelescope:2022wkp} or gravitational wave emission \cite{Maggiore:2007ulw}, but also from a theoretical perspective. In fact, given that classical GR is expected to break down at high enough energies in order to get rid of the various types of singularities the theory harbours \cite{Senovilla:2014gza}, particularly for black holes, new regular solutions could help us better understand the possibilities beyond this pessimistic scenario of ill-defined geometries that give support to our current interpretation of the universe. The theoretical exploration of nonsingular solutions is thus an important topic on its own and has motivated many works within and beyond GR, as well as in higher and lower dimensions, see e.g. \cite{Cardoso:2019rvt} for a report on the current observational status of many such proposals. In this sense, the search for new solutions in $(2+1)$ dimensions has yielded highly valuable results in a wide range of gravity-matter scenarios, 
providing novel perspectives on fundamental questions in classical and quantum gravity \cite{Carlip:1995qv}. Among this class of solutions, those obtained by Ba\~{n}ados, Teitelboim, and Zanelli (BTZ) \cite{Banados:1992gq, Banados:1992wn, Martinez:1999qi} are of particular interest and have been extensively investigated in various contexts (see, for instance, \cite{Carlip:2005zn, Sahoo:2006vz, Li:2008ws}). More recently, studies in $(2+1)$ dimensions coupled to various types of fields have managed to find regular black holes \cite{He:2017ujy, Bueno:2021krl, Estrada:2020tbz, Maluf:2022jjc}, and wormhole solutions \cite{Alencar:2021ejd, Santos:2023zrj}. 

In this work we consider $(2+1)$-dimensional scalar fields borrowing inspiration from Wheeler's notion of geon  \cite{Wheeler:1955zz, Misner:1957mt}, in the sense of self-gravitating free (scalar) fields. We thus focus on a scalar field with no potential but with a non-canonical kinetic term, bringing in that way some extra freedom to the problem. Fields of this type are known in the literature as ``k-fields'', and were originally introduced in the context of cosmology \cite{Armendariz-Picon:1999hyi,Armendariz-Picon:2000ulo}. Though the study of geons typically involves some kind of time dependence, here we consider static scenarios to investigate whether any interesting structures can be found. As we will see, nontrivial solutions do emerge. 

It is well-known that in $(3+1)$ dimensions, a static, spherically symmetric free scalar field coupled to Einstein's gravity may lead to asymptotically flat, localized solutions that, nonetheless, represent naked singularities. To obtain regular solutions one must consider oscillating fields, leading to what are known as boson stars \cite{Liebling:2012fv}, a field of great activity in the last few years \cite{Vincent:2015xta,Palenzuela:2017kcg,Olivares:2018abq,Olivares:2018abq}. Exploration of non-canonical scalar fields in $(2+1)$ dimensions can help us shed some light on whether static, nonsingular exotic solutions could be possible in the more difficult $(3+1)$ case. By considering a modified ansatz for the line element that is inspired by the $(3+1)$ static, spherical case, in this work we manage to obtain exact analytical solutions that represent asymptotically flat self-gravitating scalar objects with a clearly localized circular structure.  The compactness of these structures depends on the model parameters in a quite transparent functional form, which facilitates their analysis.  Moreover, we show that the geodesic structure of the family of solutions considered is complete, confirming in that way that such objects are nonsingular.

The paper is organized as follows. In Sec. \ref{Sec:2} we define our general setting and look for solutions of the nonlinear scalar matter source. This severely constrains the form of the scalar field Lagrangian but allows us to find exact analytical solutions. We then proceed to classify the solutions and analyse their physical properties in Sec. \ref{Sec:3}, calculating the line element and some curvature scalars. To better understand the properties of the most physically appealing solutions, we study their time-like and null geodesic structure in Sec. \ref{Sec:4} and their energy distribution in Sec. \ref{Sec:5}.  We conclude with a summary and discussion of the results in Sec.  \ref{Sec:6}.

\section{$( 2 + 1 )$-Einstein theory with nonlinear scalar field \label{Sec:2}}

Let us start by defining the action for $(2+1)$-Einstein gravity coupled to a scalar field as
\begin{equation}
\mathcal{S}=\int d^{3}x\sqrt{-g}\left(\frac{1}{2\kappa}\mathcal{R}-\frac{1}{2}L(Y)\right),\label{action1}
\end{equation} 
where $\mathcal{R}\equiv g^{\mu\nu}R_{\mu\nu}$ is the usual curvature scalar of a space-time metric $g_{\mu\nu}$ and Ricci tensor $R_{\mu\nu}$, while $L(Y)$ is an arbitrary function of the scalar field invariant $Y\equiv g^{\mu\nu}\partial_{\mu}\phi\partial_{\nu}\phi$, and the constant $\kappa = 8\pi \tilde{G}/c^{4}$ where $\tilde{G}$ stands for the Newtonian gravitational constant in two spatial dimensions, which carries units of length $[\tilde{G}] = \ell$ in natural units. 

The corresponding Einstein field equations are derived by varying the action (\ref{action1}) with respect to the metric tensor, leading to
\begin{equation}
G_{\mu\nu}=\kappa T_{\mu\nu},\label{EoMmetric}
\end{equation}
with the energy-momentum tensor given by
\begin{equation}
T_{\mu\nu}=\left(L_{Y}\partial_{\mu}\phi\partial_{\nu}\phi -\frac{1}{2}g_{\mu\nu}L\right),
\end{equation}
where $L_{Y}\equiv \frac{\partial L}{\partial Y}$. On the other hand, variation with respect to the scalar field leads to 
\begin{equation}
\frac{1}{\sqrt{-g}}\nabla_{\mu}\left(\sqrt{-g}L_{Y}g^{\mu\nu}\nabla_{\nu}\phi\right)=0.\label{EoMphi}
\end{equation}
Taking the trace of Eq.(\ref{EoMmetric}) and plugging the result back, one can rewrite this equation in the more convenient form
\begin{equation}
    R_{\mu\nu}=\kappa\left(T_{\mu\nu}-g_{\mu\nu}T\right),\label{EoMmetric2}
\end{equation}
where $T\equiv g^{\mu\nu}T_{\mu\nu}=(YL_{Y}-\frac{3}{2}L)$ is the trace of the energy-momentum tensor of the scalar field.

\subsection{Static and circularly symmetric solutions}

For static and circularly symmetric scenarios, one assumes that the scalar field profile only depends on the radial coordinate, $\phi\equiv\phi(x)$, in a coordinate system represented by $\{t, x, \theta\}$. The equation of motion (\ref{EoMphi}) can thus be expressed as
\begin{equation}
    \partial_{x}\left(\sqrt{-g}g^{xx}L_{Y}\phi_{x}\right)=0,
\end{equation}
where $\phi_{x} \equiv d\phi/dx$. In order to deal with this equation, we adopt an ansatz for the line element inspired in the choice made by Wyman \cite{Wyman:1981bd} in the $(3+1)$-dimensional case but with a modification that is crucial to bring into the $(2+1)$ scenario the philosophy behind the original choice, namely, 
\begin{equation}
ds^{2}=-e^{A(x)}dt^{2}+e^{A(x)}\frac{L_{Y}^{2}}{W^{2}(x)}dx^{2}+\frac{1}{W^{2}(x)}d\theta^{2},\label{ds2}
\end{equation}
where $A$ and $W$ are arbitrary functions of the radial coordinate $x$. This unconventional form of the line element is justified by the fact that it leads to an almost trivial scalar field equation: $\phi_{xx}=0$. Without loss of generality, this allows us to take $\phi(x)\equiv x$. As a result, the kinetic term $Y$ takes the form
\begin{equation}
Y=e^{-A}\frac{W^{2}}{L_{Y}^{2}}.\label{Y}
\end{equation}
which can only be explicitly solved once a concrete function $L(Y)$ is specified. The metric field equations (\ref{EoMmetric2}) with this choice take the form
\begin{eqnarray}
R_{\ t}^{t}&=&Y\left(\frac{A_{x}L_{Y_x}}{2L_{Y}}-\frac{A_{xx}}{2}\right)=\kappa\left(L-YL_{Y}\right),\label{Rtt} \\
R_{\ x}^{x}&=&Y\Bigg(\frac{A_{x}L_{Y_x}}{2L_{Y}}-\frac{A_{x}W_{x}}{W}-\frac{L_{Y_x}W_{x}}{L_{Y}W} \nonumber \\
&-&\frac{W_{x}^{2}}{W^{2}}-\frac{A_{xx}}{2}+\frac{W_{xx}}{W}\Bigg) = \kappa L, \label{Rxx} \\
R_{\ \theta}^{\theta}&=&Y\left(-\frac{L_{Y_x}W_{x}}{L_{Y}W}-\frac{W_{x}^{2}}{W^{2}}+\frac{W_{xx}}{W}\right) \nonumber \\
&=&\kappa\left(L-YL_{Y}\right),\label{Rthetatheta}
\end{eqnarray}
where we used the relation (\ref{Y}) and denoted $L_{Y_x} \equiv \frac{\partial L_{Y}}{\partial x}$. 

Once a scalar Lagrangian $L(Y)$ is specified, the above equations represent a nonlinear coupled system for the variables $A(x)$,  $W(x)$, and $L(x)$. A useful relation can be obtained by evaluating $R_{x}^{x} - R_{t}^{t} - R_{\theta}^{\theta}$, which leads to a first integral of the above system in the form 
\begin{equation}
\frac{A_{x}W_{x}}{W}=\kappa\frac{\left(L-2YL_{Y}\right)}{Y} \ .  \label{Yauxiliar}
\end{equation}
Another useful expression follows by rewriting Eq.(\ref{Rtt}) in the form 
\begin{equation}
A_{xx}-A_{x}\frac{L_{Y_x}}{L_{Y}}=-2\kappa\frac{\left(L-YL_{Y}\right)}{Y} \ ,
\end{equation}
which can be seen as a first-order linear ordinary differential equation for the variable $A_x$ that admits the formal solution
\begin{equation}\label{eq:EDO-1}
    A_x=L_Y\left(c_1-2\kappa\int dx\frac{(L-YL_Y)}{YL_Y}\right) \ .
\end{equation}
where $c_1$ is an integration constant. For arbitrary $L_Y$, this is an integro-differential equation for $A(x)$ coupled to $W(x)$. However, for the specific choice 
\begin{equation}
L(Y)=\lambda Y^{\alpha},\label{LYc1}
\end{equation}
with $\lambda$ a constant with suitable dimensions and $\alpha$ a dimensionless parameter\footnote{For the static scenarios we are considering, the kinetic term $Y$ is always positive. In more general settings, and in order to prevent problems if $Y$ becomes negative, one could consider a redefinition of $\alpha$ as $\alpha \to 2\tilde{\alpha}$ to force that the Lagrangian is indeed a real quantity.}, the integrand of the second term in Eq.(\ref{eq:EDO-1}) becomes a constant, allowing us to obtain an explicit solution of the form
\begin{equation}
A_{x}=-\lambda\alpha Y^{\alpha-1} \Delta_1\ ,\label{Ax}
\end{equation}
where we have defined the quantity
\begin{equation}\label{eq:Delta1}
    \Delta_1\equiv \frac{2\kappa (1-\alpha)}{\alpha}(x-x_1) \ ,
\end{equation}
and $x_1\equiv c_1 \alpha/(2\kappa(1-\alpha))$ is an integration constant. Note that since the theory is invariant under constant shifts of the scalar field, $\phi(x)\to \phi(x)+\phi_1$, and we have taken a coordinate system in which $\phi(x)=x$, the constant $x_1$ does not play any physical role and can be set to zero for simplicity. Note also that the above expressions lead to singular results in the cases of $\alpha = 1,1/2$, or $0$, so we shall first elaborate on the general case before discussing these singular ones.

Combining Eq.(\ref{Ax}) with Eq.(\ref{Yauxiliar}), we also find an important simplification, namely, 
\begin{equation}
\frac{W_{x}}{W}=\frac{\kappa(2\alpha-1)}{\alpha\Delta_1}=\frac{(2\alpha-1)}{2(1-\alpha)}\frac{1}{(x-x_1)} \ ,  \label{Yauxiliar-alpha}
\end{equation}
whose solution can be written as 
\begin{equation}\label{eq:W}
    W(x)=\frac{1}{r_0}\left(\frac{x-x_1}{x_0}\right)^{\gamma/2}
\end{equation}
where we have included an integration constant $r_0 x_0^{\gamma/2}$ for dimensional consistency, and have defined the parameter
\begin{equation}
\gamma\equiv \frac{(2\alpha-1)}{(1-\alpha)} \ ,
\end{equation}
which will play a relevant role in the classification of solutions. 
The above expression is valid as long as $\alpha\neq 0,1/2$ and $1$, which belongs to the singular cases. Using the above results in Eq.(\ref{Rthetatheta}), a bit of algebra leads to the following equation for the function $Y$:   
\begin{equation}
    Y_{x}-\frac{2\kappa}{\alpha\Delta_1}Y=\frac{\lambda\alpha \Delta_1}{(2\alpha-1)}Y^{\alpha} \ ,
\end{equation}
which turns out to be a nonlinear first-order equation of the Bernouilli type. Such equations can be linearized by the change of variable $\nu=Y^{1-\alpha}$, leading to 
\begin{equation}
\nu(x)= (x-x_1) \Delta_2\ ,
\end{equation}
where we have defined
\begin{equation}\label{eq:D2}
    \Delta_2\equiv c_2+\frac{2\kappa\lambda(1-\alpha)}{\gamma}x
\end{equation}
One can thus put $Y^{\alpha-1}=1/\nu(x)$ into Eq.(\ref{Ax}) and integrate to obtain 
\begin{equation}
e^{A(x)}=\Delta_2^{-\gamma}
\end{equation}
With all the above results, the line element for generic $\alpha\neq 0,1/2,1$ can be written as
\begin{equation} \label{eq:linele}
ds^2=-\Delta_2^{-\gamma}dt^2+\frac{(\alpha\lambda)^2 r_0^2 x_0^\gamma}{(x\Delta_2)^{(2+\gamma)}}dx^2+r_0^2\left(\frac{x_0}{x}\right)^{\gamma} d\theta^2 \ ,
\end{equation}
where we have set $x_1=0$ for simplicity. We can rewrite this line element in terms of the usual radial coordinate $r$ by identifying $r^2=1/W^2$. Using the relation (\ref{eq:W}) with $x_1=0$ for simplicity, we have that  $r^2=r_0^2 x_0^{\gamma}/x^{\gamma}$ leads to 
\begin{equation}
    \left(\frac{dr}{r}\right)^2=\left(\frac{\gamma}{2}\right)\left(\frac{dx}{x}\right)^2 \ ,
\end{equation}
and inserting this result in (\ref{eq:linele}), we get
\begin{equation} \label{eq:linele-r}
ds^2=-\Delta_2^{-\gamma}dt^2+{\sigma_0^2}\Delta_2^{-(2+\gamma)}dr^2+r^2 d\theta^2 \ ,
\end{equation}
where $\sigma_0^2\equiv (2\alpha \lambda/\gamma)^2$ is just a constant. This completes our construction of the line element of this scalar theory.

\section{Families of solutions}\label{Sec:3}

We will now proceed to classify the solutions of our model in terms of the parameter $\alpha$ that characterizes the scalar field Lagrangian or, equivalently, in terms of the exponent $\gamma$ defined in Eq.(\ref{eq:W}). The relation between these two parameters is as follows: 
\begin{itemize}
\item $\frac{1}{2}<\alpha<1$ corresponds to the interval $\gamma>0$, with $\gamma=0$ identified with $\alpha=1/2$ and $\gamma\to +\infty$ with $\alpha\to 1^-$. 
\item $\alpha>1$ leads to $-\infty<\gamma<-2$, with $\alpha\to 1^+$ corresponding to $\gamma\to -\infty$ and $\alpha\to \infty$ leading to $\gamma\to -2$.
\item $\alpha<0$ is mapped into $-2<\gamma<-1$, with $\alpha=0$ leading to $\gamma=-1$ and $\alpha\to-\infty$ to $\gamma=-2$. 
\item $0<\alpha<\frac{1}{2}$ is mapped into $-1<\gamma<0$. 
\end{itemize}
As is evident from this classification, there are three values of the parameter $\alpha$, corresponding to $\{0,1/2,1\}$, that require a separate discussion and will be considered later. We will address the features of the general case first. For this purpose, we focus on the radial dependence of the function $\Delta_2$ defined in Eq.(\ref{eq:D2}). Since (\ref{eq:W}) allows us to write
\begin{equation}
x=x_0\left(\frac{r_0}{r}\right)^{2/\gamma}     \ ,
\end{equation}
we can rewrite $\Delta_2$ as
\begin{equation}\label{eq:D2b}
    \Delta_2\equiv c_2\left[1+\frac{2\lambda\kappa(1-\alpha)x_0}{\gamma c_2}\left(\frac{r_0}{r}\right)^{2/\gamma}\right] \ .
\end{equation}
Note that the factor $c_2$ in front of this expression can be absorbed into a redefinition of the time coordinate, in a rescaling of $\sigma_0$, and in a rescaling of $\lambda\kappa x_0$ (as long as $c_2>0$, which we will assume from now on). Thus, without loss of generality, we can set  $c_2=1$ in $\Delta_2$. 

Let us now consider the radial dependence of  (\ref{eq:D2b}). We see that only when $\gamma>0$ will we have asymptotically flat solutions, which happens in the interval $1/2<\alpha<1$. These are the solutions we are mostly interested in. 
On the other hand, we see that $\lambda$ determines the sign of the second term in the square bracket. Considering the far limit, $r\to \infty$, of the $g_{tt}$ component of the metric 
in its representation (\ref{eq:linele-r}), we see that 
\begin{equation}\label{eq:gttseries}
     g_{tt}\approx -\left(1-2\lambda\kappa(1-\alpha)x_0\left(\frac{r_0}{r}\right)^{2/\gamma}+\ldots\right) \ .
\end{equation}
This expansion shows that the sign of $\lambda$ in the second term determines if the source is attractive ($\lambda>0$) or repulsive ($\lambda<0$). We can thus write $\Delta_2$ as 
\begin{equation}\label{eq:D2c}
    \Delta_2\equiv \left[1\pm\left(\frac{R_0}{r}\right)^{2/\gamma}\right] \ ,
\end{equation}
which will simplify our discussion of these asymptotically flat solutions. In this last expression we have just defined $R_0$ as 
\begin{equation}
    R_0^{2/\gamma}\equiv \frac{2|\lambda|\kappa(1-\alpha)x_0 r_0^{2/\gamma}}{\gamma c_2} \ ,
\end{equation}
and the positive sign in the bracket represents an attractive source with $\lambda>0$.



\subsection{Cases $\alpha=0,1/2$, and $1$}

Let us consider first the case $\alpha=1$, {which represents a canonical massless scalar field}. Going back to Eqs.(\ref{Ax}) and (\ref{eq:Delta1}), we see that $A_x=\lambda c_1$ can be trivially integrated to obtain $A(x)=A_0+\lambda c_1 x$. It is also easy to find that $W(x)=1/(r_0 e^{\frac{\kappa x}{c_1}})$, which allows us to define $r(x)=r_0 e^{\frac{\kappa x}{c_1}}$. Combining these results in Eq.(\ref{Y}) we get that $Y=e^{-\left(\lambda c_1+\frac{2\kappa}{ c_1}\right) x}/(\lambda r_0)^2$. In terms of the radial coordinate $r$, the line element can thus be written as 
\begin{equation}
ds^2=-\left(\frac{r}{r_0}\right)^{\lambda c_1^2/\kappa}dt^2+\left(\frac{\lambda c_1}{\kappa}\right)^2\left(\frac{r}{r_0}\right)^{\lambda c_1^2/\kappa}dr^2+r^2d\theta^2 \ ,
\end{equation}
This line element represents the exact solution for a free massless scalar field in $(2+1)$ dimensions. In $(3+1)$ dimensions, the solution for a free, massless scalar field is well known in exact form \cite{Wyman:1981bd}, and one can find an asymptotically flat solution for some choice of parameters. In that case, the solution approaches the Schwarzschild geometry far from the peak of the matter distribution, while the solution in the high-density region is formally similar to the expression found above. In fact, using the notation of \cite{Magalhaes:2022esc} (see Sec.V.A. in that paper), the internal geometry of the $(3+1)$ solution can be approximated as 
 \begin{equation}
ds^2_{3+1}=-\left(\frac{r}{2M}\right)^{\frac{8M^2}{\mu^2}}dt^2+\left(\frac{r}{2M}\right)^{\frac{8M^2}{\mu^2}}\left(\frac{4M r}{\mu^2}\right)^2dr^2+r^2d\Omega^2 \ ,
\end{equation}
where $M$ represents the asymptotic mass of the object and $\mu\ll M$ is a small mass scale. This comparison allows us to see that both cases represent naked singularities with a strong curvature and energy density divergence at $r\to 0$ (for instance, the Ricci scalar goes as $\sim r^{-(2+\lambda c_1^2/\kappa)}$). This suggests that the $(2+1)$ solutions may be seen as a rough description of the innermost regions of the $(3+1)$ configurations (at least qualitatively).
 
On the other hand, the case $\alpha=1/2$ is very peculiar because Eq.(\ref{Yauxiliar-alpha}) implies that either $W_x$ or $A_x$ must vanish. In both cases such a fact leads to $Y=0$, which generates inconsistencies in the equations. We will thus not explore this case in any further detail. 

Finally, when $\alpha=0$, the original matter action reduces to a cosmological constant-type term, resulting in the well-known BTZ black hole solution \cite{Banados:1992gq, Banados:1992wn}.


\section{Asymptotically flat solutions}

Let us now focus on the line element (\ref{eq:linele-r}) with $\Delta_2$ defined as in (\ref{eq:D2c}), and with $\gamma>0$. For positive $\lambda$, the line element becomes
\begin{equation}
    ds^{2}=-\frac{1}{\left(1+\left(\frac{R_0}{r}\right)^{\frac{2}{\gamma}}\right)^\gamma}dt^{2}+\frac{\sigma_0^2  dr^2}{\left(1+\left(\frac{R_0}{r}\right)^{\frac{2}{\gamma}}\right)^{2+\gamma}}  + r^{2}d\theta^{2} \ , \label{ds2schild2+}
\end{equation}
and represents a horizonless space-time with a delicate point at $r\to 0$. 
For negative $\lambda$, the line element is
\begin{equation}
    ds^{2}=-\frac{1}{\left(1-\left(\frac{R_0}{r}\right)^{\frac{2}{\gamma}}\right)^\gamma}dt^{2}+\frac{\sigma_0^2  dr^2}{\left(1-\left(\frac{R_0}{r}\right)^{\frac{2}{\gamma}}\right)^{2+\gamma}}  + r^{2}d\theta^{2} \ , \label{ds2schild2}
\end{equation}
there are no horizons either, but the delicate point arises as $r\to R_0>0$. In both cases, the problems cannot be avoided by a redefinition of the radial coordinate because they affect also the $g_{tt}$ component. A look at the Ricci and Kretschmann curvature scalars leads to 
\begin{eqnarray}
R&=& \pm\frac{2 (\gamma +4) \left(1\pm\left(\frac{R_0}{r}\right)^{\frac{2}{\gamma} }\right)^{\gamma +1} \left(\frac{R_0}{r}\right)^{\frac{2}{\gamma} }}{\gamma \sigma_0^2 r^2 } \\
K&=& \frac{4 \left(3 \gamma ^2+8 \gamma +8\right) \left(1\pm\left(\frac{R_0}{r}\right)^{\frac{2}{\gamma} }\right)^{2 (\gamma +1)} \left(\frac{R_0}{r}\right)^{\frac{4}{\gamma} }}{\gamma ^2 \sigma_0^4 r^4 } \ , \nonumber
\end{eqnarray}
where the $\pm$ sign corresponds to the sign of $\lambda$. When $\lambda>0$, curvature scalars  diverge as $r\to 0$, pointing towards a curvature singularity caused by the concentration of energy at the center (recall that $\lambda>0$ represents an attractive field). For negative $\lambda$, instead, the curvature vanishes as $r\to R_0$ and we gain no new information about what may be happening in that region to generate a divergence in the metric. To deepen into this aspect and try to unveil what is really going on in that region, we must study the behavior of geodesics.\\

\subsection{Geodesics}\label{Sec:4}

We will now explore whether the solutions found above represent singular or nonsingular spacetimes from the perspective of their geodesic structure. For this purpose, we must determine if the affine parameter is defined over the whole real line (complete geodesics) or it in can only cover a portion of it (incomplete geodesics). A space-time with any nonzero number of incomplete geodesics is regarded as singular. This is so because incomplete null geodesics imply that information (light rays) can be created and/or destroyed, while incomplete time-like geodesics imply that observers can be created and/or destroyed, which is physically unacceptable. This is the key notion behind the theorems proving the existence of space-time singularities within GR, see e.g. \cite{Senovilla:2014gza} for a discussion of this topic.

The Lagrangian from which the geodesic equations can be obtained can be written as 
\begin{widetext}
\begin{equation}
    L=\frac{1}{2}\left(\frac{ds}{d\tau}\right)^{2}=\frac{1}{2}\left[-\left(1\pm\left(\frac{R_0}{r}\right)^{\frac{2}{\gamma}}\right)^{-\gamma}\dot{t}^{2}+\sigma_0^2 \left(1\pm\left(\frac{R_0}{r}\right)^{\frac{2}{\gamma}}\right)^{-2-\gamma}\dot{r}^{2}+r^{2}\dot{\theta}^{2}\right],
\end{equation}
\end{widetext}where $\tau$ is an affine parameter (the proper time for time-like observers), and the overdot denotes differentiation with respect to it. Taking into account the symmetries of the Lagrangian, i.e., its static and invariant nature under rotations, we find the presence of two conserved quantities, as given by
\begin{eqnarray}
    E&=&-\frac{\partial L}{\partial\dot{t}}=\left(1\pm\left(\frac{R_0}{r}\right)^{\frac{2}{\gamma}}\right)^{-\gamma}\dot{t},\label{energy} \\
    J&=&\frac{\partial L}{\partial \dot{\theta}}=r^{2}\dot{\theta},\label{momentum}
\end{eqnarray}
where $E$ and $J$  denote the energy and angular momentum per unit mass of the particle, respectively. As usual, we can normalize the 4-velocity $U^{\mu}$ to one,  so that
\begin{equation}
    U^{\mu}U_{\mu}=\left(\frac{ds}{d\tau}\right)^{2}\equiv\epsilon=\pm1,0,\label{velocity}
\end{equation}
where the parameter $\epsilon$ characterizes the type of geodesics we are dealing with: time-like ($\epsilon=-1$), space-like ($\epsilon=1$), or null ($\epsilon=0$). 

Combining Eqs.(\ref{energy}), (\ref{momentum}), and (\ref{velocity}), we can conveniently write the geodesic equation as
\begin{equation}\label{eq:geodesics}
\frac{\sigma_0^2 \left(\frac{dr}{d\tau}\right)^2}{\left(1\pm\left(\frac{R_0}{r}\right)^{2/\gamma}\right)^{2+2\gamma}}=E^2-\frac{1}{\left(1\pm\left(\frac{R_0}{r}\right)^{2/\gamma}\right)^{\gamma}}\left(\frac{J^2}{r^2}-\epsilon\right) \ . 
\end{equation}
Since the left-hand side of this equation must be positive by construction, if $\epsilon=-1$ or $J\neq 0$ the domain of $r(\tau)$ must be restricted to the region 
\begin{equation}
E^2\ge \frac{1}{\left(1\pm\left(\frac{R_0}{r}\right)^{2/\gamma}\right)^{\gamma}}\left(\frac{J^2}{r^2}-\epsilon\right)
\end{equation}
This means that any massive particle ($\epsilon=-1$) or massless particle with angular momentum ($J\neq 0$) that moves inwards in the radial direction, will eventually reach a minimum, $r_{m}\ge 0$ if $\lambda>0$ and $r_{m}\ge R_0$ if $\lambda<0$, at which the equality above is satisfied. The motion then must continue towards increasing values of $r$ (the particle moves away after reaching the closest radial distance), thus guaranteeing the completeness of all such geodesics. 

Considering now radial null geodesics, i.e., those with $\epsilon=0$ and $J=0$, then Eq.(\ref{eq:geodesics}) can be written in the simpler form
\begin{equation}\label{eq:radialnull}
\frac{d\hat{r}}{\left(1\pm\left(\frac{1}{\hat{r}}\right)^{2/\gamma}\right)^{1+\gamma}} =\pm d\hat{\tau}\ , 
\end{equation}
where we have defined the dimensionless variables $\hat{r}\equiv r/R_0$ and also $\hat{\tau}={E \tau}/{R_0\sigma_0}$. This equation can be formally integrated for arbitrary $\gamma$, yielding the result
\begin{equation}\label{eq:2F1}
 \, _2F_1\left(-\frac{\gamma }{2},\gamma +1;1-\frac{\gamma }{2};\mp \hat{r}^{-2/\gamma }\right) \hat{r}=\pm \hat \tau+\beta \ ,
\end{equation}
where $\beta$ is an integration constant and $_2F_1$ a hypergeometric function. Note that the $\pm$ sign in $_2F_1$ is associated with the sign of $\lambda$, while on the right-hand side it represents if the geodesic is outgoing ($+$) or ingoing ($-$). In the far limit, where $\hat{r}\to \infty$, we can approximate this hypergeometric function by $\ _2F_1\left(-\frac{\gamma }{2},\gamma +1;1-\frac{\gamma }{2};0\right)=1$, which leads to $\hat{r}\mp \hat \tau=\alpha$ and represents the usual straight lines of light rays in asymptotically flat geometries. In the opposite limit, we need to split the discussion because for $\lambda>0$, the limit corresponds to $\hat{r}\to 0$, while for $\lambda<0$, we have $\hat{r}\to 1$. 

The expansion around $\hat{r}\to 0$ can be easily derived from (\ref{eq:radialnull}) by approximating the left-hand side as $ \hat{r}^{2+2/\gamma}d\hat{r}$. By direct integration we find that radial null geodesics in this region behave as 
\begin{equation}
    \frac{\hat{r}^{3+2/\gamma}}{3+2/\gamma}\approx \pm \hat \tau+\tilde{\beta} \ , 
\end{equation}
which implies that they reach $\hat{r}\to 0$ in finite affine time, confirming that this space-time is singular, as we had guessed from the curvature scalars.  

Let us now focus on the case with $\lambda<0$. When $\hat{r}\to 1$, one can show that the dominant term of the solution takes the form
\begin{equation}
 -\frac{1}{2}\left(\frac{\gamma/2}{\hat{r}-1}\right)^\gamma\approx \pm \hat \tau+\beta
\end{equation}
and diverges as $\hat{r}\to 1$. This means that the affine parameter always diverges as the minimal circumference $r=R_0$ is approached, implying that all these geodesics are complete. Thus, the circumference $r=R_0$ represents a boundary of the manifold and cannot be reached in finite affine time. Together with the completeness of the other geodesics discussed above, this results into a nonsingular space-time despite the divergence of the metric functions in that region. Note that this situation has been found before in the literature within other gravitational settings, see e.g. \cite{Bambi:2015zch}.

\begin{figure*}[t!]
  \centering
  \includegraphics[width=0.45\textwidth]{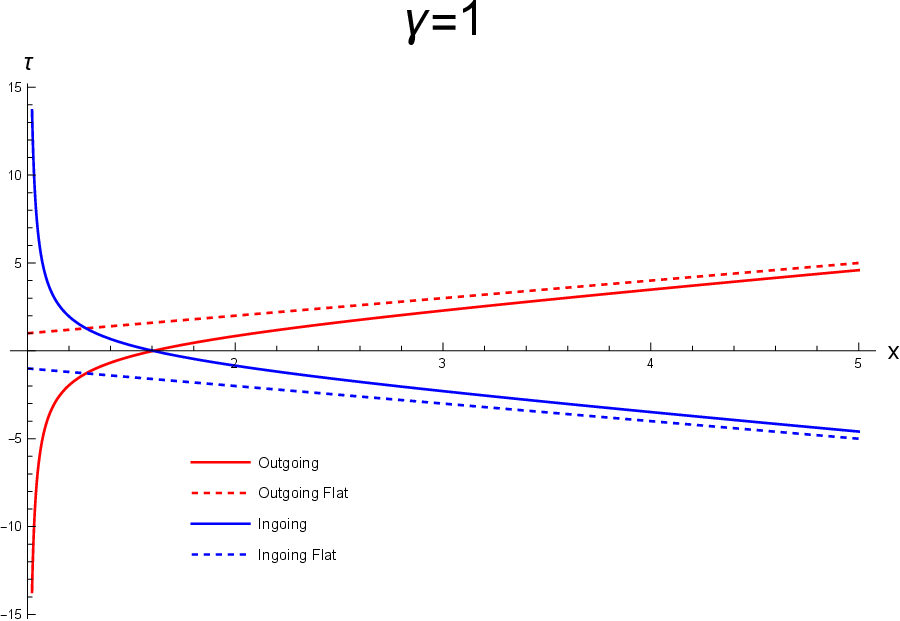}
  \hspace{0.05\textwidth}
  \includegraphics[width=0.45\textwidth]{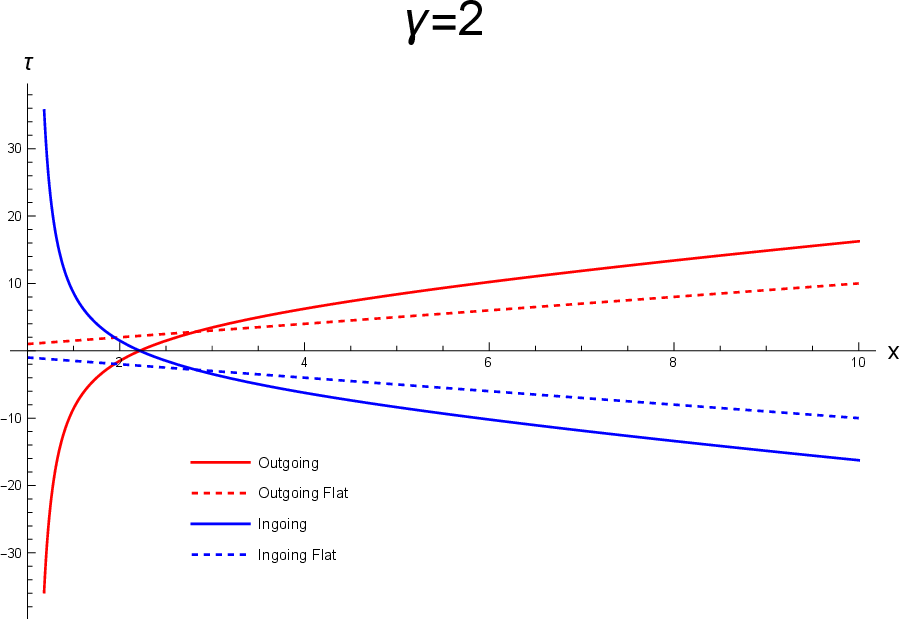}\\ 
  \includegraphics[width=0.45\textwidth]{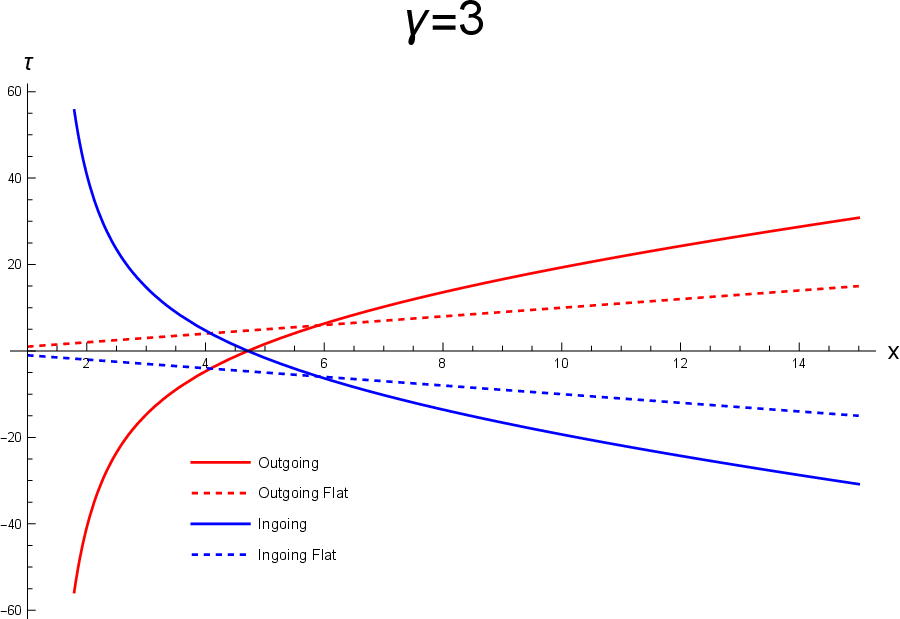}
  \hspace{0.05\textwidth}
  \includegraphics[width=0.45\textwidth]{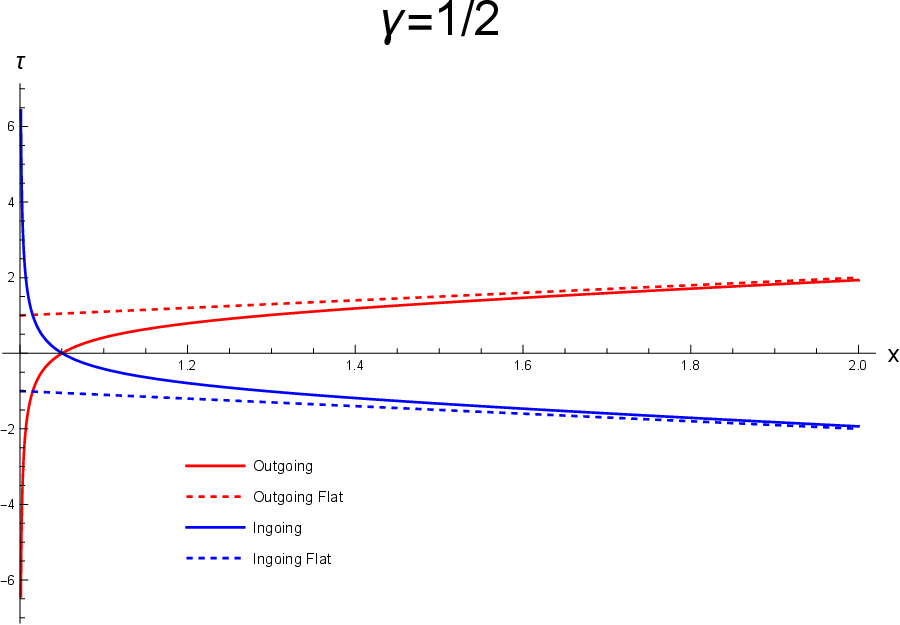}\\
    \includegraphics[width=0.45\textwidth]{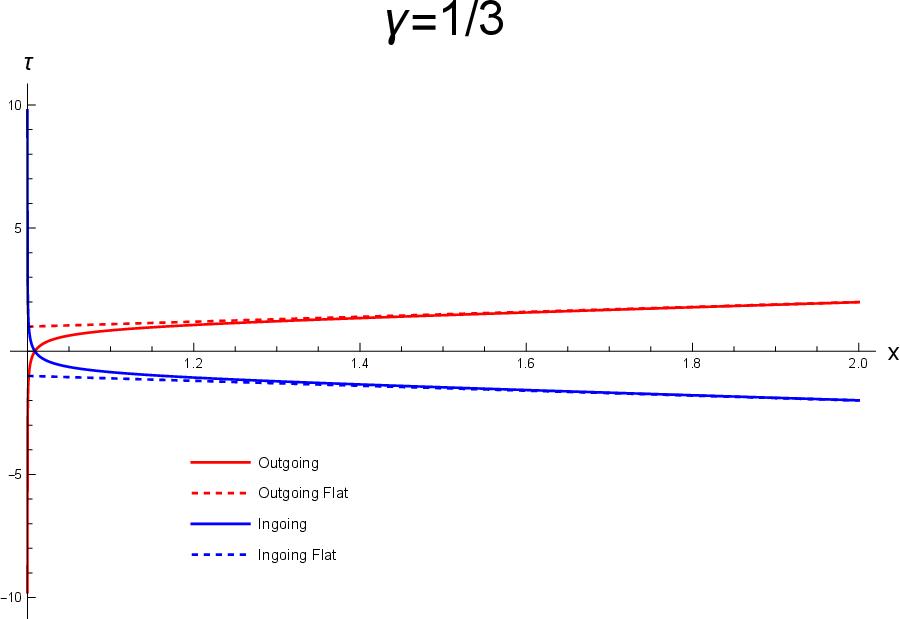}
  \hspace{0.05\textwidth}
  \includegraphics[width=0.45\textwidth]{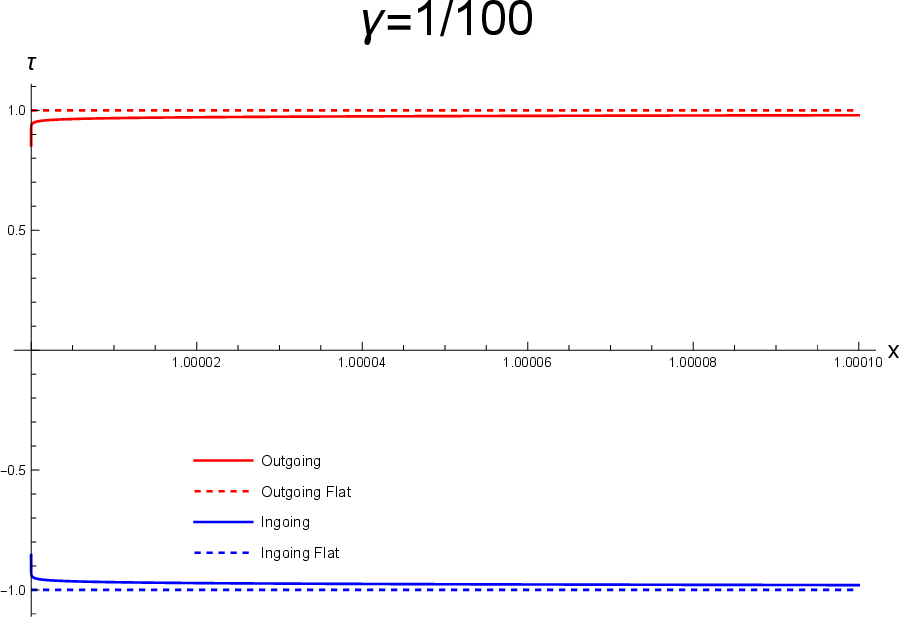}\\ 
  \caption{Representation of the (normalized) affine parameter $\frac{E }{R_0\sigma_0^{1/2} } \tau$ as a function of the radial coordinate $\hat{r}$ for ingoing (blue) and outgoing (red) geodesics. The dashed lines depict the trajectories $\pm \hat{r}$ that represent the Minkowskian geodesics (for illustration). Note that for $0<\gamma<1$ the convergence to the Minkowskian value is very fast, being fastest in the limit $\gamma\to 0$. The divergence of the affine parameter as $\hat{r}\to 1$ shows that this region is a boundary of the manifold that cannot be reached by any observer or light signal.  \label{fig:geodesics}}
\end{figure*}

The left-hand side of Eq.(\ref{eq:2F1}) admits a representation in terms of elementary functions for some values of the parameter $\gamma$. Some examples for $\lambda<0$ are as follows:
\begin{itemize}
\item $\gamma=1 \Rightarrow \ \ \  \frac{\hat{r}}{2-2 \hat{r}^2}+\hat{r}+\frac{3}{4} \log \frac{(\hat{r}-1)}{(\hat{r}+1)}$ 
\item $\gamma=2 \Rightarrow \ \ \ \hat{r}+\frac{5-6 \hat{r}}{2 (\hat{r}-1)^2}+3 \log (\hat{r}-1)$ 
\item $\gamma=3 \Rightarrow \ \ \  \frac{1}{16} \left(\frac{-693 \hat{r}^{5/3}+144 \hat{r}^{7/3}+16 \hat{r}^3-315 \sqrt[3]{\hat{r}}+840 \hat{r}}{\left(\hat{r}^{2/3}-1\right)^3}\right.$ \\ $ \ \ \ \ \ \ \ \ \ \left.-315 \left[\tanh ^{-1}\left(\sqrt[3]{\hat{r}}\right)+\frac{i\pi}{2}\right]\right)$ 
\end{itemize}
The representation of radial null geodesics for these and other values of the parameter $\gamma$ appears in Fig. \ref{fig:geodesics}, where their completeness, as given by the affine parameter $\tau$ going to $\pm \infty$ at both ends of the coordinate $\hat{r}$, i.e., $\hat{r} \to \infty$ and $\hat{r} \to 1$, is evident.

\section{Energy density distribution}\label{Sec:5}
Let us now focus on how the energy density is distributed in the solutions studied above. From Eq.(\ref{Y}) and a little algebra using the line element (\ref{ds2schild2}), we see that the kinetic term $Y$ can be written as 
\begin{equation}
Y=Y_0 \frac{\left(1\pm\hat{r}^{-2/\gamma}\right)^{2+\gamma}}{\hat{r}^{\frac{2(2+\gamma)}{\gamma}}} \ ,
\end{equation}
where $Y_0\equiv \left(\lambda\alpha R_0\right)^{-\frac{2(2+\gamma)}{\gamma}}$ is an irrelevant constant factor. 

For the singular solutions corresponding to $\lambda>0$, it is easy to see that this kinetic energy density diverges when $\hat r\to 0$ as $Y/Y_0\approx 1/r^{4(2+\gamma)/\gamma}$, which provides further evidence about its pathological nature.

\begin{figure*}[t!]
  \centering
  \includegraphics[width=0.45\textwidth]{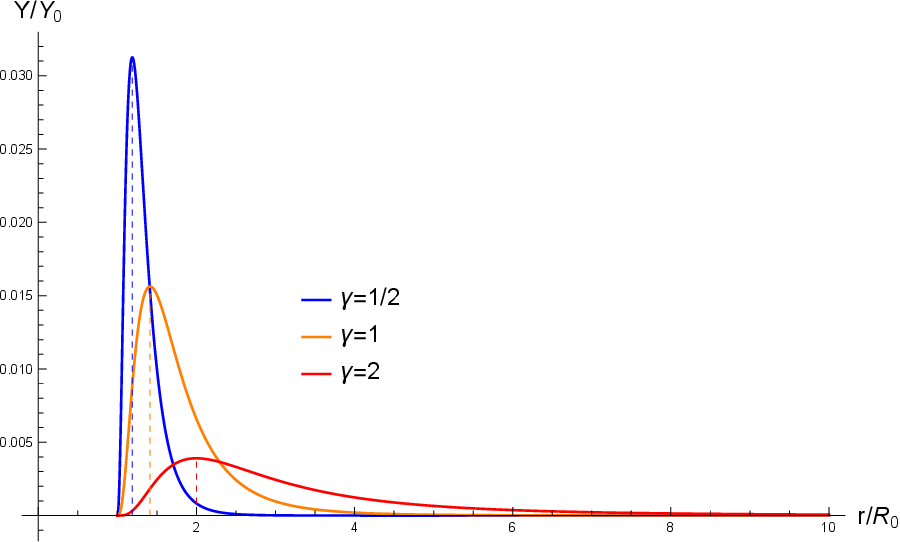}
  \caption{ Representation of the kinetic term $Y=g^{\mu\nu}\partial_\mu\phi\partial_\nu\phi$ for the cases $\gamma=1/2,1,$ and $2$ (recall that $\alpha=(1+\gamma)/(2+\gamma)$) when $\lambda<0$. Vertical dashed lines indicate the location of the maximum. The localized nature of these solutions is evident. Note that the smaller the value of $\gamma$, the higher the peak and the more compact the structure. \label{fig:Y2D}}
\end{figure*}

\begin{figure*}[t!]
  \centering
  \includegraphics[width=0.45\textwidth]{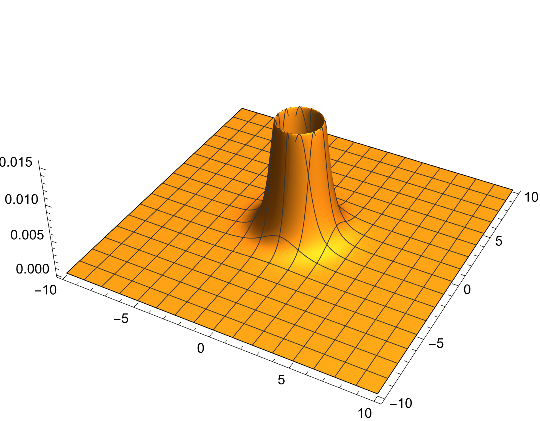}
   \hspace{0.05\textwidth}
  \includegraphics[width=0.45\textwidth]{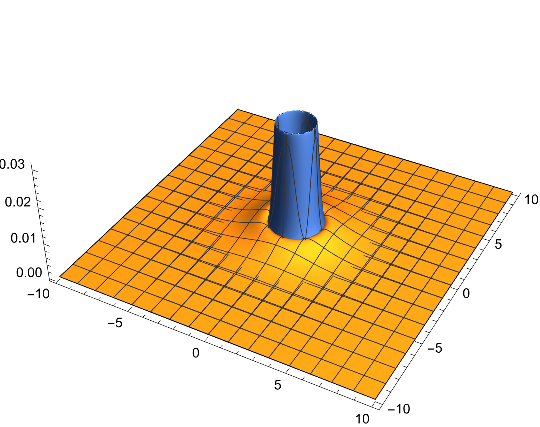}
  \caption{Left: three-dimensional  representation of the kinetic term $Y=g^{\mu\nu}\partial_\mu\phi\partial_\nu\phi$ when $\lambda<0$ for the case $\gamma=1$. Right: same representation for $\gamma=1/2$ (blue) and $\gamma=2$ (orange). Note how the more compact solution $\gamma=1/2$  is always hidden by the $\gamma=2$ one except at the innermost region. The different amplitudes of the maxima are also evident in this plot. \label{fig:Y3D1}} 
\end{figure*}

On the contrary, for $\lambda<0$ the energy density goes to zero both at infinity and at the minimal circumference $\hat{r}=1$, both of which represent boundaries of the manifold. One thus expects the existence of a maximum located somewhere in between these two asymptotic regions. An elementary calculation indicates that $Y_{\hat{r}}$ vanishes at $\hat{r}=1$, at infinity, and at $\hat{r}_M=2^{\gamma/2}$, where the kinetic term takes the maximum value $Y=Y_0 2^{-2(2+\gamma)}$. Therefore, our nonsingular solutions represent localized energy distributions with a maximum around the circumference of radius $r=R_02^{\gamma/2}$.  A representation of the amplitude of the kinetic term $Y$ on the plane is provided in Figs. \ref{fig:Y2D} and \ref{fig:Y3D1}. One can verify by direct calculation that the total energy of the system is finite (integrating the scalar action from $\hat{r}=1$ to infinity) for all $\gamma>0$.

\section{Summary and conclusion\label{Sec:6}}

In this work we have studied $(2+1)$-dimensional Einstein gravity coupled to a static, nonlinear scalar field with a purely kinetic term and circular symmetry. The search for analytical solutions led us to consider a family of power-law models with a Lagrangian density of the form given in Eq.(\ref{LYc1}). After classifying the various branches of solutions, we focused on the case $\gamma>0$ (equivalently $1/2<\alpha < 1$) and showed that the resulting geometries are determined by the line element (\ref{ds2schild2}), which represents asymptotically flat spaces. We showed that when the parameter $\lambda$ that sets the amplitude of the scalar Lagrangian is positive, we have an attractive source, whereas for negative $\lambda$ we have a repulsive source. All solutions with  $\lambda>0$ represent naked singularities (divergent curvatures and energy density, and incomplete geodesics), whereas for $\lambda<0$ all solutions are regular and nonsingular. 

Though in the $\lambda<0$ case the $g_{tt}$ and $g_{rr}$ components of the metric diverge at $r=R_0>0$, we found that curvature invariants vanish at that location. Furthermore, we showed that the circumference $r=R_0$ represents a boundary of the manifold, as all radial null geodesics take an infinite affine time to reach there. Time-like geodesics and null rays with nonzero angular momentum never reach this boundary and have an $r>R_0$ as their minimal radial coordinate. The analysis of the kinetic term of the scalar field shows that these geometries are generated by localized volcano-like lumps of energy with maximum amplitude at $r=2^{\gamma/2}R_0$, remaining positive everywhere and vanishing only at $r=R_0$ and at infinity (see Figs. \ref{fig:Y2D} and \ref{fig:Y3D1}). 

In our view, the most relevant result of this paper is the discovery of exact analytical solutions that represent nonsingular static compact scalar objects in an asymptotically flat geometry. These localized structures are possible thanks to the exotic (non-canonical) dynamics of the scalar field, and the fact that they generate an inner boundary of radius $R_0$ is a surprise that could have not been anticipated a priori. In practical terms, this boundary and its neighborhood act like a region of repulsive forces (because geodesics bounce) that prevent the collapse of the energy distribution and regularise its maximum amplitude. Even though this kind of exotic matter source has repulsive gravitational properties, it is worth exploring its stability and potential interactions with other sources to better understand alternative singularity avoidance mechanisms. 

If analogous structures could be found in $(3+1)$-dimensional extensions of this model, there could be important implications for the astrophysics of compact objects and dark matter/energy models. In particular, boson stars are regarded as spherical distributions of scalar matter with peak density at the center. Our analysis puts forward that nonsingular compact objects without a center do exist within GR. This means that contrary to the standard approach, one should look for new solutions of self-gravitating scalar fields with boundary conditions which are not defined at a center, since the later may not exist.  These and other related questions are currently under study.

\section*{Acknowledgments}

The authors thank the Funda\c{c}\~{a}o Cearense de Apoio ao Desenvolvimento Cient\'{i}fico e Tecnol\'{o}gico (FUNCAP), the Coordena\c{c}\~{a}o de Aperfei\c{c}oamento de Pessoal de N\'{i}vel Superior (CAPES), and the Conselho Nacional de Desenvolvimento Cient\'{i}fico e Tecnol\'{o}gico (CNPq), Grant no. 200879/2022-7 (RVM). 
This work is also supported by the Spanish Agencia Estatal de  Investigaci\'on (grants PID2020-116567GB-C21 and PID2022-138607NB-I00, funded by MCIN/AEI/10.13039/501100011033, FEDER, UE, and ERDF A way of making Europe) and by the project PROMETEO/2020/079 (Generalitat Valenciana). R. V. Maluf thanks the Department of Theoretical Physics \& IFIC  of the University of Valencia - CSIC, and the Department of Theoretical Physics and IPARCOS of the Complutense University of Madrid, for the kind hospitality during the elaboration of this work. This article is based upon work from COST Action CA21136, supported by COST (European Cooperation in Science and Technology).



\begin{thebibliography}{99}


\bibitem{Stephani:2003tm}
H.~Stephani, D.~Kramer, M.~A.~H.~MacCallum, C.~Hoenselaers and E.~Herlt,
``Exact solutions of Einstein's field equations'', 
Cambridge Univ. Press, 2003.

\bibitem{Penrose:1969pc}
R.~Penrose,
Riv. Nuovo Cim. \textbf{1} (1969) 252.

\bibitem{Carter:1971zc}
B.~Carter,
Phys. Rev. Lett. \textbf{26} (1971) 331.

\bibitem{Kerr:1963ud}
R.~P.~Kerr,
Phys. Rev. Lett. \textbf{11} (1963) 237.


\bibitem{Newman:1965my}
E.~T.~Newman, {\it et al.},
J. Math. Phys. \textbf{6} (1965) 918.

\bibitem{EventHorizonTelescope:2019dse}
K.~Akiyama \textit{et al.} [Event Horizon Telescope],
Astrophys. J. Lett. \textbf{875} (2019) L1.

\bibitem{EventHorizonTelescope:2022wkp}
K.~Akiyama \textit{et al.} [Event Horizon Telescope],
Astrophys. J. Lett. \textbf{930} (2022)  L12.

\bibitem{Maggiore:2007ulw}
M.~Maggiore,
``Gravitational Waves. Vol. 1: Theory and Experiments,''
Oxford University Press, 2007.


\bibitem{Senovilla:2014gza}
J.~M.~M.~Senovilla and D.~Garfinkle,
Class. Quant. Grav. \textbf{32} (2015) 124008.


\bibitem{Cardoso:2019rvt}
V.~Cardoso and P.~Pani,
Living Rev. Rel. \textbf{22} (2019)  4.

\bibitem{Carlip:1995qv}
S.~Carlip,
Class. Quant. Grav. \textbf{12} (1995) 2853.

\bibitem{Banados:1992gq}
M.~Banados, M.~Henneaux, C.~Teitelboim and J.~Zanelli,
Phys. Rev. D \textbf{48} (1993) 1506
[erratum: Phys. Rev. D \textbf{88} (2013) 069902].

\bibitem{Banados:1992wn}
M.~Banados, C.~Teitelboim and J.~Zanelli,
Phys. Rev. Lett. \textbf{69} (1992) 1849.

\bibitem{Martinez:1999qi}
C.~Martinez, C.~Teitelboim and J.~Zanelli,
Phys. Rev. D \textbf{61} (2000) 104013.


\bibitem{Carlip:2005zn}
S.~Carlip,
Class. Quant. Grav. \textbf{22} (2005) R85.

\bibitem{Sahoo:2006vz}
B.~Sahoo and A.~Sen,
JHEP \textbf{07} (2006) 008.

\bibitem{Li:2008ws}
R.~Li and J.~R.~Ren,
Phys. Lett. B \textbf{661} (2008) 370.

\bibitem{He:2017ujy}
Y.~He and M.~S.~Ma,
Phys. Lett. B \textbf{774} (2017) 229.


\bibitem{Bueno:2021krl}
P.~Bueno, P.~A.~Cano, J.~Moreno and G.~van der Velde,
Phys. Rev. D \textbf{104} (2021)  L021501.


\bibitem{Estrada:2020tbz}
M.~Estrada and F.~Tello-Ortiz,
EPL, \textbf{135} (2021) 20001.

\bibitem{Maluf:2022jjc}
R.~V.~Maluf, C.~R.~Muniz, A.~C.~L.~Santos and M.~Estrada,
Phys. Lett. B \textbf{835} (2022) 137581.

\bibitem{Alencar:2021ejd}
G.~Alencar, V.~B.~Bezerra and C.~R.~Muniz,
Eur. Phys. J. C \textbf{81} (2021) 924.

\bibitem{Santos:2023zrj}
A.~C.~L.~Santos, C.~R.~Muniz and R.~V.~Maluf,
JCAP \textbf{09} (2023), 022.

\bibitem{Wheeler:1955zz}
J.~A.~Wheeler,
Phys. Rev. \textbf{97} (1955) 511.

\bibitem{Misner:1957mt}
C.~W.~Misner and J.~A.~Wheeler,
Annals Phys. \textbf{2} (1957) 525.

\bibitem{Armendariz-Picon:1999hyi}
C.~Armendariz-Picon, T.~Damour and V.~F.~Mukhanov,
Phys. Lett. B \textbf{458} (1999) 209.

\bibitem{Armendariz-Picon:2000ulo}
C.~Armendariz-Picon, V.~F.~Mukhanov and P.~J.~Steinhardt,
Phys. Rev. D \textbf{63} (2001) 103510.

\bibitem{Liebling:2012fv}
S.~L.~Liebling and C.~Palenzuela,
Living Rev. Rel. \textbf{26} (2023)  1.

\bibitem{Vincent:2015xta}
F.~H.~Vincent, Z.~Meliani, P.~Grandclement, E.~Gourgoulhon and O.~Straub,
Class. Quant. Grav. \textbf{33} (2016) 105015.

\bibitem{Palenzuela:2017kcg}
C.~Palenzuela, P.~Pani, M.~Bezares, V.~Cardoso, L.~Lehner and S.~Liebling,
Phys. Rev. D \textbf{96} (2017) 104058.

\bibitem{Olivares:2018abq}
H.~Olivares, {\it et. al.}, 
Mon. Not. Roy. Astron. Soc. \textbf{497} (2020)  521.

\bibitem{Wyman:1981bd}
M.~Wyman,
Phys. Rev. D \textbf{24} (1981) 839.

\bibitem{Magalhaes:2022esc}
R.~B.~Magalh\~aes, L.~C.~B.~Crispino and G.~J.~Olmo,
Phys. Rev. D \textbf{105} (2022) 064007

\bibitem{Bambi:2015zch}
C.~Bambi, A.~Cardenas-Avendano, G.~J.~Olmo and D.~Rubiera-Garcia,
Phys. Rev. D \textbf{93} (2016) 064016.


\end{thebibliography}
\end{document}